\input{aipcheck}

\documentclass[
    ,final            
  ]
  {aipproc}

\layoutstyle{6x9}

\usepackage{bm}
\newcommand{\br}{\bm{r}}

\begin{document}

\title{Halos in a deformed Relativistic Hartree-Bogoliubov theory in continuum}


\author{Lulu Li}{
  address={Institute of Applied Physics and Computational Mathematics, Beijing 100094, China}
}

\author{Jie Meng}{
  address={State Key Laboratory of Nuclear Physics and Technology, School of Physics, 
              Peking University, Beijing 100871, China}
  ,altaddress={School of Physics and Nuclear Energy Engineering, 
              Beihang University, Beijing 100191, China}
  ,altaddress={Department of Physics, University of Stellenbosch, Stellenbosch, South Africa}
}

\author{P. Ring}{
  address={Physikdepartment, Technische Universit\"at M\"unchen,
              85748 Garching, Germany}
  ,altaddress={State Key Laboratory of Nuclear Physics and Technology, School of Physics, 
              Peking University, Beijing 100871, China}
}

\author{En-Guang Zhao}{
  address={State Key Laboratory of Theoretical Physics,
              Institute of Theoretical Physics, Chinese Academy of Sciences,
              Beijing 100190, China}
  ,altaddress={Center of Theoretical Nuclear Physics, National Laboratory
              of Heavy Ion Accelerator, Lanzhou 730000, China} 
}

\author{Shan-Gui Zhou}{
  address={State Key Laboratory of Theoretical Physics,
              Institute of Theoretical Physics, Chinese Academy of Sciences,
              Beijing 100190, China}
  ,altaddress={Center of Theoretical Nuclear Physics, National Laboratory
              of Heavy Ion Accelerator, Lanzhou 730000, China} 
}

\begin{abstract}
In this contribution we present some recent results about neutron halos in deformed nuclei. 
A deformed relativistic Hartree-Bogoliubov theory in continuum 
has been developed and the halo phenomenon in deformed weakly bound nuclei is investigated. 
These weakly bound quantum systems present interesting examples 
for the study of the interdependence between the deformation of 
the core and the particles in the halo. 
Magnesium and neon isotopes are studied and detailed results are 
presented for the deformed neutron-rich and weakly bound nuclei $^{42}$Mg. 
The core of this nucleus is prolate, but the halo has a slightly oblate shape. 
This indicates a decoupling of the halo orbitals from the deformation of the core. 
The generic conditions for the existence of halos in deformed nuclei and 
for the occurrence of this decoupling effect are discussed.
\end{abstract}

\maketitle


\section{\label{sec:intro}Introduction}

Halo is one of the most interesting exotic nuclear phenomena.
In halo nuclei, the extremely weakly binding property results in many new features, 
e.g., the coupling between bound states and the continuum due to pairing 
correlations and very extended spatial density distributions. 
Therefore one must consider properly the asymptotic behavior of nuclear densities
at large distance $r$ from the center and treat in a self consistent
way the discrete bound states, the continuum and the coupling between
them in order to give a proper theoretical description of the halo
phenomenon.
This can be achieved by solving the non-relativistic
Hartree-Fock-Bogoliubov (HFB)~\cite{Bulgac1980_nucl-th9907088,
Dobaczewski1984_NPA422-103, Dobaczewski1996_PRC53-2809} or
the relativistic Hartree Bogoliubov
(RHB)~\cite{Meng1996_PRL77-3963, Poschl1997_PRL79-3841, Lalazissis1998_PLB418-7,
Meng1998_NPA635-3, Meng2006_PPNP57-470} equations in coordinate ($r$) space which
can fully take into account the mean-field effects of the coupling to the continuum.

For spherical nuclei, relativistic Hartree-Bogoliubov theories
in coordinate space have been developed~\cite{Meng1996_PRL77-3963, 
Poschl1997_PRL79-3841, Lalazissis1998_PLB418-7, Meng1998_NPA635-3, Meng2006_PPNP57-470}.
With the relativistic continuum Hartree-Bogoliubov (RCHB) theory~\cite{Meng1998_NPA635-3, 
Meng2006_PPNP57-470}, 
properties of the halo nucleus $^{11}$Li has been reproduced quite well~\cite{Meng1996_PRL77-3963}
and the prediction of giant halos in light and medium-heavy nuclei was made~\cite{Meng1998_PRL80-460,
Meng2002_PRC65-041302R, Zhang2003_SciChinaG46-632}.
The RCHB theory has been generalized to treat the odd particle system~\cite{Meng1998_PLB419-1}
and combined with the Glauber model, 
the charge-changing cross sections for C, N, O and F isotopes on a carbon target
have been reproduced well~\cite{Meng2002_PLB532-209}.

Since most open shell nuclei are deformed, the interplay between deformation 
and weak binding raises interesting questions, such as whether or not 
there exist halos in deformed nuclei and, if yes, what are their new features.
In order to consider properly the asymptotic behavior of nuclear densities
at large $r$ and to make the numerical procedure less complicated, 
the Woods-Saxon basis has been proposed in Ref.~\cite{Zhou2003_PRC68-034323}
as a reconciler between the harmonic oscillator basis and the integration
in coordinate space. 
Over the past years, lots of efforts have been made to develop a deformed relativistic
Hartree theory~\cite{Zhou2006_AIPCP865-90} and a deformed relativistic Hartree Bogoliubov
theory in continuum (the DefRHBC theory)~\cite{Meng2003_NPA722-C366, Zhou2008_ISPUN2007, 
Zhou2010_PRC82-011301R, Zhou2011_JPCS312-092067, Li2012_PRC85-024312}. 
In order to describe the exotic nuclear structure in unstable odd-$A$ 
or odd-odd nuclei, the DefRHBC theory has been extended to incorporate 
the blocking effect due to the odd nucleon(s)~\cite{Li2012_CPL29-042101}. 
The deformed relativistic Hartree-Bogoliubov theory in continuum with 
the density-dependent meson-nucleon couplings is developed 
recently~\cite{Chen2012_PRC85-067301}.

The halo phenomenon in deformed nuclei has been investigated with the DefRHBC theory~\cite{Zhou2011_JPCS312-092067, Li2012_PRC85-024312}. 
In some deformed neutron-rich and weakly bound nuclei, e.g., $^{42,44}$Mg,
a decoupling of the halo orbitals from the deformation of the core has been predicted. 
In this contribution, the results of the DefRHBC theory on the study of 
deformed halo nuclei will be presented. 

\section{\label{sec:formalism} Theoretical framework}

The Dirac Hartree Bogoliubov (RHB) equation for the nucleons reads~\cite{Kucharek1991_ZPA339-23},
\begin{eqnarray}
 \int d^3 \bm{r}'
 \left(
  \begin{array}{cc}
   h_D
   - \lambda &
   \Delta
   \\
  -\Delta^*
   & -h_D
   + \lambda \\
  \end{array}
 \right)
 \left(
  { U_{k}
  \atop V_{k}
   }
 \right)
 & = &
 E_{k}
  \left(
   { U_{k}
   \atop V_{k}
    }
  \right)
 ,
 \label{eq:RHB0}
\end{eqnarray}
where $E_{k}$ is the quasiparticle energy, $\lambda$ is the chemical potential,
and $h_D$ is the Dirac Hamiltonian,
\begin{equation}
 h_D(\bm{r}, \bm{r}') =
  \bm{\alpha} \cdot \bm{p} + V(\bm{r}) + \beta (M + S(\bm{r})).
\label{eq:Dirac0}
\end{equation}
with scalar and vector potentials
\begin{eqnarray}
S(\bm{r}) & = & g_\sigma \sigma(\bm{r}), \label{eq:vaspot}\\
V(\bm{r}) & = & g_\omega \omega^0(\bm{r}) +g_\rho \tau_3 \rho^0(\bm{r})
                    +e \displaystyle\frac{1-\tau_3}{2} A^0(\bm{r}) .
\label{eq:vavpot}
\end{eqnarray}
In the particle-particle (pp) channel, we use a density dependent  zero range force,
\begin{equation}
 V^\mathrm{pp}(\br_1,\br_2) =  V_0 \frac{1}{2}(1-P^\sigma)\delta( \mathbf{r}_1 - \mathbf{r}_2 )
   \left(1-\frac{\rho(\br_1)}{\rho_\mathrm{sat}}\right).
 \label{eq:pairing_force}
\end{equation}
$\frac12(1-P^\sigma)$ projects onto spin $S=0$ component in the pairing field.
The pairing potential then reads,
\begin{equation}
 \Delta(\br)=V_0(1-\rho(\br)/\rho_{\rm sat})\kappa(\br) ,
\end{equation}
and we need only the local part of the pairing tensor
\begin{equation}
 \kappa(\br)= \sum_{k>0} V_{k}^\dagger(\bm{r})U^{}_{k}(\bm{r}) .
\label{E12}
\end{equation}

For axially deformed nuclei with the spatial reflection symmetry, we
expand the potentials $S(\bm{r})$ and $V(\bm{r})$ in Eq.~(\ref{eq:Dirac0})
and various densities in terms of the Legendre
polynomials~\cite{Price1987_PRC36-354},
\begin{equation}
 f(\bm{r})   = \sum_\lambda f_\lambda({r}) P_\lambda(\cos\theta),\
 \lambda = 0,2,4,\cdots
 ,
 \label{eq:expansion}
\end{equation}
with an explicit definition of $f_\lambda({r})$.

The quasiparticle wave functions $U_k$ and $V_k$ in Eq.~(\ref{eq:RHB0}) are  
expanded in the Woods-Saxon basis~\cite{Zhou2003_PRC68-034323}:
\begin{eqnarray}
 U_{k} (\br{s} p)
 & = & \displaystyle
 \sum_{n\kappa} u^{(m)}_{k,(n\kappa)}     \varphi_{n\kappa m}(\br{s} p),
 \label{eq:Uexpansion0} \\
 V_{k} (\br{s} p)
 & = & \displaystyle
 \sum_{n\kappa} v^{(m)}_{k,(n\kappa)} \bar\varphi_{n\kappa m}(\br{s} p).
\label{eq:Vexpansion0}
\end{eqnarray}
$\bar\varphi_{n\kappa m}(\br{s} p)$ is the time reversal state of $\varphi_{n\kappa
m}(\br{s} p)$. 
Because of the axial symmetry the $z$-component $m$ of
the angular momentum $j$ is a conserved quantum number and the RHB Hamiltonian can
be decomposed into blocks characterized by $m$ and parity $\pi$. For
each $m^\pi$-block, solving the RHB equation (\ref{eq:RHB0}) is
equivalent to the diagonalization of the matrix
\begin{equation}
 \left( \begin{array}{cc}
  {\cal A}-\lambda & {\cal B} \\
  {\cal B^\dag} & -{\cal A}^\ast+\lambda \\
 \end{array} \right)
 \left(
  { {\cal U}_k
    \atop
    {\cal V}_k
  }
 \right)
 = E_k
 \left(
  { {\cal U}_k
    \atop
    {\cal V}_k
  }
 \right),
 \label{eq:RHB1}
\end{equation}
where
\begin{equation}
 {\cal U}_k = \left(u^{(m)}_{k,(n\kappa)}\right),\
 {\cal V}_k = \left(v^{(m)}_{k,(n\kappa)}\right),
\end{equation}
and
\begin{eqnarray}
 {\cal A}
 & = &
 \left( h^{(m)}_{D(n\kappa)(n'\kappa')} \right)
 = 
 \left( \langle n\kappa m|h_D|n'\kappa',m\rangle \right) ,
 \\
 {\cal B}
 & = &
 \left( \Delta^{(m)}_{(n\kappa)(n'\kappa)} \right)~
 = 
 \left( \langle n\kappa m |\Delta| \overline{n'\kappa',m} \rangle \right).
\label{eq:pairing_matrix}
\end{eqnarray}
Further details can be found in the appendixes of Ref.~\cite{Li2012_PRC85-024312}.

\section{\label{sec:results}Results and discussions}

We next present some results from the DefRHBC theory 
by taking magnesium isotopes as 
examples and discuss in details results for the deformed neutron-rich and 
weakly bound nucleus $^{42}$Mg~\cite{Zhou2010_PRC82-011301R, Zhou2011_JPCS312-092067, 
Li2012_PRC85-024312}. 

Magnesium isotopes have been studied extensively in 
Refs.~\cite{Zhou2010_PRC82-011301R, Li2012_PRC85-024312} with the deformed
relativistic Hartree-Bogoliubov theory in continuum and the parameter sets
NL3~\cite{Lalazissis1997_PRC55-540} and PK1~\cite{Long2004_PRC69-034319}. 
In the calculations based on the parameter set PK1, $^{42}$Mg is the last
bound nucleus in Mg isotopes~\cite{Li2012_PRC85-024312}.

\begin{figure}
\includegraphics[width=0.4\columnwidth]{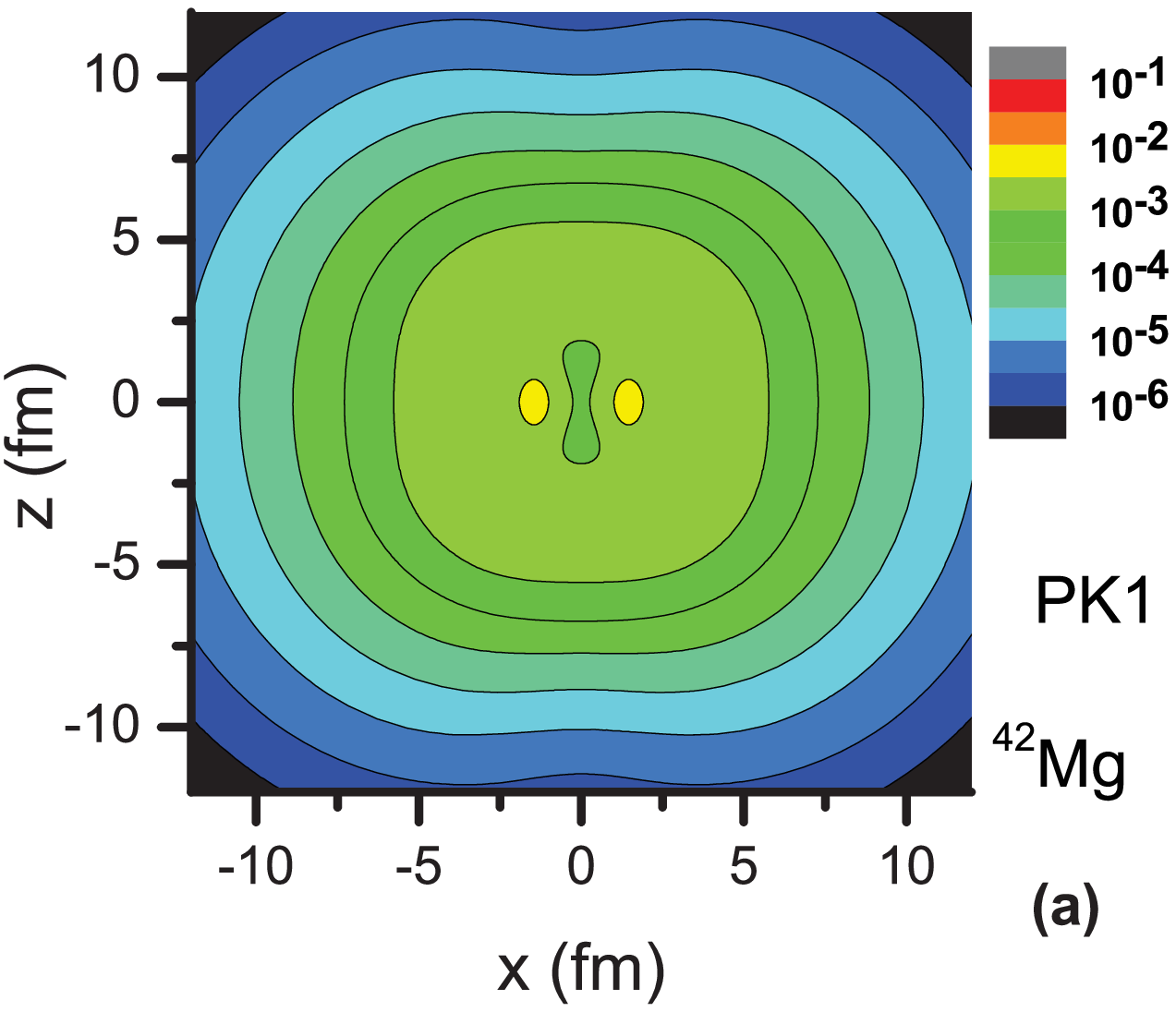}
~~~~
\includegraphics[width=0.4\columnwidth]{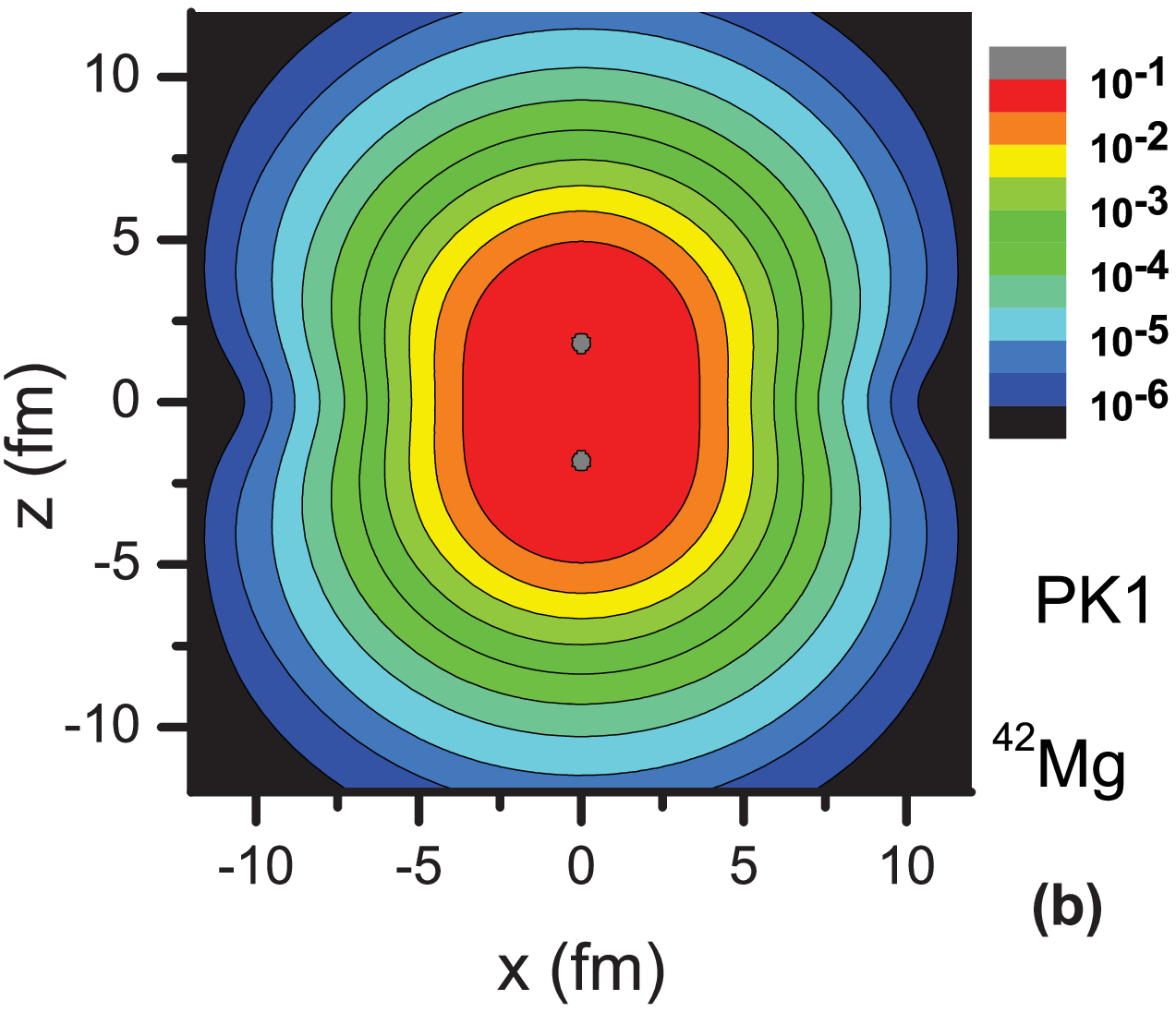}
\caption{(Color online) Density distributions of the ground state of $^{42}$Mg with the
$z$ axis as the symmetry axis: (a) the neutron halo, and (b) the neutron core.
Taken from Ref.~\cite{Li2012_PRC85-024312}.
}
\label{fig:Mg42_pro_2d}
\end{figure}

It was found in Ref.~\cite{Li2012_PRC85-024312} that 
the ground state of $^{42}$Mg is well deformed with 
a quadrupole deformation $\beta \approx 0.41$,
and a very small two neutron separation energy $S_{2n} \approx 0.22$ MeV.
In the tail part, the neutron density
extends more along the direction perpendicular to the symmetry axis.
The density distribution is decomposed into contributions of the oblate 
``halo'' and of the prolate ``core'' in Fig.~\ref{fig:Mg42_pro_2d}.
The density distribution of this weakly bound nucleus has a very long tail
in the direction perpendicular to the symmetry axis
which indicates the prolate nucleus $^{42}$Mg has an oblate halo and
there is a decoupling between the deformations of the core and the halo.

\begin{figure}
\includegraphics[width=0.5\columnwidth]{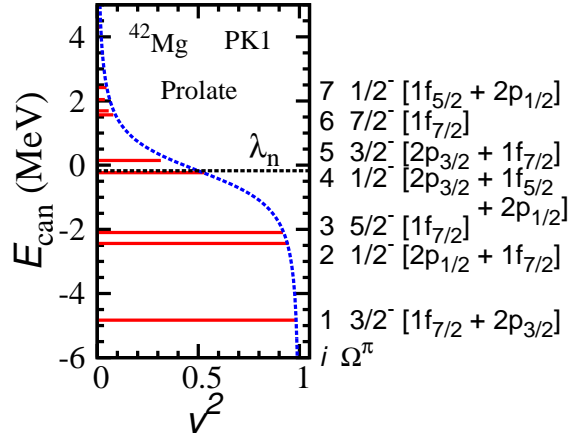}
\caption{(Color online) 
Single neutron levels of ground state of $^{42}$Mg in the canonical basis
as a function of the occupation probability $v^2$.
The blue dashed line corresponds to the BCS-formula with an average pairing gap.
Taken from Ref.~\cite{Li2012_PRC85-024312}.
}
\label{fig:Mg42_pro_lev}
\end{figure}

The single particle spectrum around the Fermi level for the ground state of
$^{42}$Mg is shown in Fig.~\ref{fig:Mg42_pro_lev}~\cite{Li2012_PRC85-024312}.
The good quantum numbers of each single particle state are also shown.
The occupation probabilities $v^2$ in the canonical basis have BCS-form~\cite{Ring1980}
and are given by the length of the horizontal lines in Fig.~\ref{fig:Mg42_pro_lev}.
The levels close to the threshold are labeled by the number $i$ according
to their energies, and their conserved quantum number
$\Omega^\pi$ as well as the main spherical components are given at the right hand side.
The neutron Fermi level is within the $pf$ shell and most of the single
particle levels have negative parities.
Since the chemical potential $\lambda_n$ is close to the continuum, orbitals
above the threshold have noticeable occupations due to the pairing correlations.
The single neutron levels of $^{42}$Mg can be divided into two parts, the
deeply bound levels ($\varepsilon_{\rm can} < -2$ MeV) corresponding
to the ``core'', and the remaining weakly bound levels close to the
threshold ($\varepsilon_{\rm can} > -0.3$ MeV) and in the continuum
corresponding to the ``halo''.

As discussed in Refs.~\cite{Zhou2010_PRC82-011301R, Zhou2011_JPCS312-092067, Li2012_PRC85-024312},
the shape of the halo originates from the intrinsic structure of the weakly
bound or continuum orbitals.
By examining the neutron density distribution, we learned that 
for the ground state of $^{42}$Mg, the halo is mainly
formed by level 4 and level 5.
Note that the angular distribution of $ |Y_{10} (\theta, \phi)|^2 \propto
\cos^2 \theta$ with a projection of the orbital angular momentum on the
symmetry axis $\Lambda = 0$ is prolate and that of
$ |Y_{1\pm 1} (\theta, \phi)|^2 \propto \sin^2 \theta$ with
$\Lambda = 1$ is oblate.
For level 4 ($\Omega^\pi = 1/2^-$), $\Lambda$ could be 0 or 1 since
the third component of the total spin is $1/2$.
However, it turns out that the $\Lambda=1$ component dominates which results in
an oblate shape.
For level 5, since the third component of the total spin is $3/2$, $\Lambda$
can only be 1, which corresponds to an oblate shape too.
Therefore in $^{42}$Mg the shape of the halo is oblate and decouples from the prolate core.

The decoupling between the deformations of the core and the halo
may manifest itself by some new experimental observables, e.g.,
the double-hump shape of longitudinal momentum distribution in 
single-particle removal reactions and new dipole modes, etc.
In particular, a combination of the experimental method proposed in 
Ref.~\cite{Navin1998_PRL81-5089} and the theoretical approach developed 
in Ref.~\cite{Sakharuk1999_PRC61-014609} would be useful in the study of 
longitudinal momentum distribution in single-particle removal reactions 
with deformed halo nuclei as projectiles.
The shape decoupling effects may also has some influence on the 
sub-barrier capture process in heavy ion 
collisions~\cite{Sargsyan2011_2012}.

For odd particle system, the formation and the size of a halo depend
strongly on the interplay among the odd-even effects, continuum and pairing effects,
deformation effects, etc. Some progress on this topic has been made recently~\cite{Li_in-prep}.

\section{Summary}

We present recent progresses of the development of 
a deformed relativistic Hartree-Bogoliubov theory in continuum (DefRHBC) 
and the study of neutron halo in deformed nuclei.
In very neutron-rich deformed nuclei $^{42,44}$Mg, pronounced deformed
neutron halos were predicted. 
The halo is formed by several orbitals close to the threshold. 
These orbitals have large components of low $\ell$-values
and feel therefore only a small centrifugal barrier. 
Although their cores are prolately deformed, 
the deformation of the halos is slightly oblate. 
This implies a decoupling between the shapes of the core and the halo. 
The mechanism is investigated by studying the details of the neutron densities 
for core and halo, the single particle levels in the canonical basis, 
and the decomposition of the halo orbitals. 
It was concluded that the existence and the deformation of a possible neutron halo 
depends essentially on the quantum numbers of the main components of the single particle orbits
in the vicinity of the Fermi surface.

\begin{theacknowledgments}
This work has been supported by the Major State Basic
Research Development Program of China (973 Program:
``New physics and technology at the limits of nuclear
stability''), NSF of China
(10875157, 10975100, 10979066, 11105005, 11175002, 11175252,
11121403, and 11120101005), the Knowledge Innovation Project of CAS 
(KJCX2-EW-N01 and KJCX2-YW-N32), 
and the DFG cluster of excellence \textquotedblleft Origin and Structure of the
Universe\textquotedblright\ (www.universe-cluster.de). 
The results described in this Letter were obtained
on the ScGrid of Supercomputing Center, Computer
Network Information Center of Chinese Academy of
Sciences.
\end{theacknowledgments}


\end{document}